\documentclass[pra,twocolumn,showpacs,superscriptaddress,eqsecnum]{revtex4}
\usepackage{amsfonts,amssymb,bm,graphicx,times}
\newcommand{\Tr}{\mathop{\mathrm{Tr}}\nolimits}
\newcommand{\op}[1]{\hat{#1}}

\begin{document}

\title{Quantum light depolarization: the phase-space perspective}

\author{A. B. Klimov}
\affiliation{Departamento de F\'{\i}sica,
Universidad de Guadalajara, 44420~Guadalajara,
Jalisco, Mexico}

\author{J. L. Romero}
\affiliation{Departamento de F\'{\i}sica,
Universidad de Guadalajara, 44420~Guadalajara,
Jalisco, Mexico}

\author{L. L. S\'anchez-Soto}
\affiliation{Departamento de \'{O}ptica,
Facultad de F\'{\i}sica, Universidad Complutense,
28040~Madrid, Spain}

\author{A. Messina}
\affiliation{Dipartimento di Scienze Fisiche ed Astronomiche,
Universit\'a di Palermo, via Archirafi 36, 90123 Palermo, Italy}

\author{A. Napoli}
\affiliation{Dipartimento di Scienze Fisiche ed Astronomiche,
Universit\'a di Palermo, via Archirafi 36, 90123
  Palermo, Italy} \date{\today}

\begin{abstract}
  Quantum light depolarization is handled through a master 
  equation obtained by coupling dispersively the field to a 
  randomly distributed atomic reservoir. This master equation 
  is solved by transforming it into a quasiprobability distribution 
  in phase space and the quasiclassical limit is investigated.
\end{abstract}

\maketitle

\section{Introduction}

Polarization of light is a key concept that has deserved a lot of
attention over the years. Apart from its fundamental significance, 
it is also of interest in several active technological fields. 
In many of these applications it is crucial to determine the 
decrease of the degree of polarization experienced by the light 
when traversing an optical system: we refer to this as 
depolarization~\cite{Brosseau1998}.

In classical optics, this depolarization is ascribed, broadly speaking,
either to birefringence (as it usually happens, e.g., in optical
fibers~\cite{Crosignani1986,Matera1995,Karlsson1998,VanWiggeren1999,
Savory2001,Nascimento2005,Mecozzi2007}) or to scattering by randomly
distributed particles~\cite{Hulst1981,Barber1990,Ishimaru1999}. In
both cases the net result is an effective anisotropy that leads to 
a decorrelation of the phases of the electric field vector.

In quantum optics, a sensible approach to deal with this decorrelation
is through the notion of decoherence, by which we loosely understand
the appearance of irreversible and uncontrollable quantum correlations
when a system interacts with its environment~\cite{Zurek2003}.
Usually,  decoherence is accompanied by dissipation, i.e., a
net exchange of  energy with the environment. However, we are
interested in the case  of pure decoherence (also known as
dephasing), for which the process  of energy dissipation is
negligible.  Models in which the populations do not change,
while the coherences are strongly decaying, are at
hand~\cite{Gardiner2004,Weiss1999,Breuer2002}.  Since a good
knowledge  of dephasing is of utmost importance (prominently
for quantum information  processing, where operations completely
rely on the presence of  coherence), these models have been
successfully applied to dephasing  in systems such as quantum
dots~\cite{Takagahara1999,Uskov2000,Krummheuer2002,Pazy2002}, Josephson
junctions~\cite{Makhlin1999,Nakamura1999,vanderWal2000,Makhlin2004},
or general quantum registers~\cite{Reina2002}, to cite only a few
relevant examples.

In the modern parlance of quantum information it is usual to call
depolarizing channel a decoherence induced by an unbiased noise
generating bit-flip and phase-flip errors~\cite{Nielsen2000}.  
While this terminology fits well with abstract qubits, whenever 
light is concerned there is an extra essential ingredient to be 
taken into account: quantum polarization has a quite natural 
su(2) invariance that leads to a structure of invariant
subspaces~\cite{Karassiov1993}. On physical grounds, we argue 
that this structure must be preserved in the evolution, which 
makes previous approaches to fail in this case.

The purpose of this paper is twofold. First, we wish to provide 
a simple model that goes around this drawback and gives a picture 
of the mechanisms involved in depolarization. The main idea is to 
couple the field dispersively to a randomly distributed atomic 
reservoir: the resulting master equation has a quite appealing
structure that complies with our requirements of su(2) invariance. 
Our second goal is to solve this equation in phase space, presenting 
then a representation of the depolarizing dynamics that makes an
easy contact with the classical one on the Poincar\'e sphere.

\section{su(2) polarization structure}
\label{su2}

We assume a monochromatic plane wave propagating in the $z$ direction,
whose electric field lies in the $xy$ plane. Under these conditions,
we have a two-mode field that can be described by two complex
amplitude operators. They are denoted by $\op{a}_{+}$ and
$\op{a}_{-}$, when using the basis of circular (right and left)
polarizations, which we shall employ in this paper. The commutation
relations of these operators are standard:
\begin{equation}
  \label{bccr}
  [ \op{a}_{\lambda}, \op{a}_{\lambda^\prime}^\dagger ] =
  \delta_{\lambda \lambda^\prime} \op{\openone} \, ,
  \qquad
  \lambda, \lambda^\prime \in \{+, - \} \, .
\end{equation}
The Stokes operators are then defined as the quantum counterparts of
the classical variables, namely~\cite{Jauch1976,Collett1970,Chirkin1993,
Alodjants1999,Luis2000}
\begin{eqnarray}
  \label{Stokop}
  & \op{S}_0 = \op{a}^\dagger_{+} \op{a}_{+} +
  \op{a}^\dagger_{-} \op{a}_{-} \, ,
  \qquad
  \op{S}_1 = \hat{a}^\dagger_{+} \op{a}_{-} +
  \op{a}^\dagger_{-} \op{a}_{+} \, , & 
  \nonumber \\
  & & \\
  & \op{S}_2 = i ( \op{a}_{+} \op{a}^\dagger_{-} -
  \op{a}^\dagger_{+} \op{a}_{-} ) \, ,
  \qquad
  \op{S}_3 = \op{a}^\dagger_{+} \op{a}_{+} -
  \op{a}^\dagger_{-} \op{a}_{-} \, , &
  \nonumber
\end{eqnarray}
and their mean values are precisely the Stokes parameters $(\langle
\op{S}_0 \rangle, \langle \op{\mathbf{S}} \rangle )$, where
$\op{\mathbf{S}} = (\op{S}_1, \op{S}_2, \op{S}_3)$. Using the relation
(\ref{bccr}), one immediately gets that the Stokes operators satisfy
the commutation relations
\begin{equation}
  \label{crsu2}
  [\hat{\mathbf{S}}, \hat{S}_0] = 0 \, ,
  \qquad
  [ \hat{S}_1, \hat{S}_2] = 2 i \hat{S}_3 \, ,
\end{equation}
and cyclic permutations. Since $\op{S}_{0}$ is just the operator
representing the total number of photon, the first equation in
(\ref{crsu2}) means that $\op{\mathbf{S}}$ is measurable in
photon-counting experiments and we can treat each energy manifold
separately. To bring out this point more clearly, it is advantageous
to relabel the standard two-mode Fock basis in the form
\begin{equation}
  \label{invsub}
  | N, k \rangle = | k \rangle_{+} \otimes
  |N - k \rangle_{-} \, ,
  \qquad
  k = 0, 1, \ldots,  N \, .
\end{equation}
For each fixed value of the number of photons $N$, these states span
an invariant subspace of dimension $N+1$ and the operators
$\op{\mathbf{S}}$ act therein according to
\begin{eqnarray}
  \op{S}_{+} \, | N, k \rangle & = &
  2 \sqrt{( k +1) ( N- k)} \, |N, k + 1 \rangle \, ,
  \nonumber \\
  \op{S}_{-} \, | N, k \rangle & = &
  2 \sqrt{k (N - k + 1)} | N, k - 1 \rangle \, , \\
  \op{S}_{3} \, | N, k \rangle & = &
  2 ( k - N/2 ) | N, k \rangle \, , \nonumber
\end{eqnarray}
where $\op{S}_{\pm} = \op{S}_{1} \pm i \hat{S}_{2}$ are raising and
lowering operators.  These invariant subspaces will play a key role 
in the following.

The quantities $\hat{\mathbf{S}}$ are the generators of the su(2)
algebra.  The noncommutability of these operators precludes the
simultaneous exact measurement of their physical quantities. Among
other consequences, this implies that no field state (leaving aside
the two-mode vacuum) can have definite nonfluctuating values of all
the Stokes operators simultaneously.  This is expressed by the
uncertainty relation
\begin{equation}
  (\Delta \op{\mathbf{S}} )^2 =
  (\Delta \op{S}_1)^2 + (\Delta \op{S}_2)^2
  + (\Delta \op{S}_3)^2 \geq 2 \langle
  \op{S}_0 \rangle \, .
\end{equation}
Contrary to what happens in classical optics, the electric vector 
of a monochromatic quantum field never describes a definite ellipse.

Finally, we recall that standard definition of the degree of
polarization reads as
\begin{equation}
  \label{doP}
  \mathbb{P} = \frac{| \langle \op{\mathbf{S}} \rangle |}
  {\langle \op{S}_{0} \rangle} =
  \frac{\sqrt{\langle \op{S}_{1}\rangle^2 +
  \langle \op{S}_{2}\rangle^2 +
  \langle \op{S}_{3}\rangle^2}} 
  {\langle \op{S}_{0} \rangle }  \, .
\end{equation}
Note that $\mathbb{P}$ depends exclusively on the first moments 
of the Stokes operators. Higher order moments can be crucial 
for a full understanding of some
phenomena~~\cite{Klimov2005,Refregier2005,Refregier2006}, but
(\ref{doP}) is more than enough for our purposes here.

\section{Depolarizing dynamics in a nonresonant atomic medium}

Pure dephasing dynamics occurs when the system Hamiltonian
$H_{\mathrm{sys}}$ commutes (at least, approximately) with the
system-reservoir interaction or for initial states that evolve very
slowly under the dynamics governed by $H_{\mathrm{sys}}$ on the time
scale of decoherence processes~\cite{Alicki2004}. Various theoretical
scenarios have been proposed to deal with this: apart from minor
details, all of them can be modelled by a scattering process in which
a reservoir quantum can be absorbed or emitted, but the number of
excitations in the system is unchanged~\cite{Gardiner2004}. These
models indeed preserve the invariant subspaces, but produce no
thermalization: they merely maintain the occupation probabilities,
while erasing all coherences. Therefore, they fail to describe light
depolarization because depolarization not only preserves the invariant
subspaces, but the steady state in each one of them must be a diagonal
state~\cite{Blum1996}. We believe that these conditions are essential
to ensure a correct description of any depolarizing process.

To solve these difficulties we take another route: we assume that the
field mode of frequency $\omega$ propagates through a material medium
made of a collection of two-level atoms with transition frequencies
$\omega_{a}$. In consequence, our basic system is represented by (in
units $\hbar = 1$ that we shall use throughout all this paper)
\begin{equation}
  \op{H}_{\mathrm{sys}} = \op{H}_{\mathrm{f}} +
  \op{H}_{\mathrm{a}} +  \op{H}_{\mathrm{fa}}  \, ,
\end{equation}
where
\begin{eqnarray}
  \label{btla}
  \op{H}_{\mathrm{f}} & = & \sum_{\lambda=\pm} \omega \,
  \op{a}_{\lambda}^\dagger \op{a}_{\lambda} \, ,  \nonumber \\
  \op{H}_{\mathrm{a}} & = & \sum_{a} \frac{1}{2} \omega_{a} \,
  \op{\sigma}_{a}^{z} \, , \\
  \op{H}_{\mathrm{fa}} & = & \sum_a \sum_{\lambda=\pm}
  ( g_{a \lambda} \, \op{\sigma}_{a}^{-} \op{a}_{\lambda}^\dagger +
  g_{a \lambda}^\ast \, \op{\sigma}_{a}^{+} \op{a}_{\lambda} ) \, .
  \nonumber
\end{eqnarray}
Here, as is normal in practice, we have described each individual atom
in terms of the standard Pauli operators. The sum over $a$ runs over
all the atoms and the interaction $\op{H}_{\mathrm{fa}}$ is written in
the dipolar and rotating-wave approximations. In addition, the atoms
are randomly distributed so the coupling constants $g_{a \lambda}$
have random phases: we thus write them as $g_{a \lambda} = |g_a | e^{i
  \lambda \varphi_a/2} $, for $\lambda = \pm$. This random-phase
approximation~\cite{Ashcroft1976} is used in many areas of physics to
estimate response functions and it works properly in the
long-wavelength limit, a hypothesis implicit in the form of the
Hamiltonian~(\ref{btla}).

It is a well-known fact that atoms decay irreversibly. This is
usually assigned to their interaction with the continuum of modes
of an additional thermal electromagnetic environment. When we take
this point into account, the density matrix for the system
(\ref{btla}) evolves according to
\begin{eqnarray}
  \label{ime}
  \dot{\op{\varrho}}_{\mathrm{sys}}(t) & =
  & - i [\op{H}_{\mathrm{sys}}, \op{\varrho}_{\mathrm{sys}}]  
  \nonumber \\
  & + & \sum_{a} \frac{\gamma_{a}}{2} \{ (\bar{n}_{a} + 1)
  \mathcal{L} [\op{\sigma}_{a}^{-} ] \, \op{\varrho}_{\mathrm{sys}} +
  \bar{n}_a \mathcal{L} [ \op{\sigma}_{a}^{+} ] \, 
  \op{\varrho}_\mathrm{sys} \} \, , \nonumber \\
  & &
\end{eqnarray}
where $\mathcal{L}[ \op{C} ]$ are Lynblad superoperators~\cite{Lindblad1976}
\begin{equation}
  \mathcal{L} [ \op{C} ] \, \op{\varrho} = 
  2 \op{C} \op{\varrho} \op{C}^\dagger
  - \{ \op{C}^\dagger \op{C}, \op{\varrho} \} \, ,
\end{equation}
and $\gamma_a$ is the decay constant of the $a$th atom due to its
coupling to the thermal environment with $\bar{n}_a$ excitations. 
Note that, by introducing a different bosonic reservoir for each 
atom, we avoid the occurrence of any collective effect.  Whereas 
a collective dephasing can introduce remarkable dynamical
changes~\cite{Yu2003,Yu2004,Banaszek2004,Ball2004,Ball2005,Ban2006},
we expect it plays no relevant role in understanding the
quasiclassical limit of depolarizing dynamics we are 
interested here~\cite{Yu2006}.

The properties of a random medium are well reproduced when $\bar{n}_a
\gg 1$~\cite{Kofman2001,Jakob2004}. In this high-temperature limit
the effect of spontaneous emission can be disregarded in comparison
with stimulated processes. Emission into the reservoir and
absorption from the reservoir therefore become identical; i. e., 
they balance each other in the stationary state: the emission and
absorption processes depend solely on the initial population of the
atomic state. Consequently, the steady-state reduced density operator
is approximately given by a mixture of equally populated atomic states
and the density matrix of the $a$th atom becomes thus diagonal
($\op{\varrho}_a = \frac{1}{2} \op{\openone}$).

If $\Delta_{a} = \omega_{a} -\omega $ is the detuning, we  consider the
far off-resonant regime
\begin{equation}
  |g_{a}| \ll | \Delta_{a} | \, .
\end{equation}
In this limit we can adiabatically eliminate the atomic variables
and obtain a master equation that, after averaging over the random
phases, reads as (see  Ref.~\cite{Klimov2006} for technical details 
on the derivation)
\begin{equation}
  \label{mesb}
  \dot{\op{\varrho}} = - i \omega  [\op{S}_{0}, \op{\varrho} ] + 2
  \gamma \, \mathcal{L}[ \op{S}_{0} ] \,  \op{\varrho}  +
  \gamma \, \mathcal{L}[ \op{S}_{+} ] \, \op{\varrho} +
  \gamma \, \mathcal{L}[ \op{S}_{-} ] \, \op{\varrho} \, ,
\end{equation}
where $\op{\varrho} (t) = \Tr_{\mathrm{at}}[ \op{\varrho}_{\mathrm{sys}}(t)]$
is the reduced density operator for the field mode and the decoherence
rate $\gamma $ is
\begin{equation}
  \label{gam}
  \gamma = \sum_{a} \frac{|g_{a}|^4}
  {\gamma_{a} \, \Delta^2 \, \bar{n}_{a}} \,.
\end{equation}
The master equation (\ref{mesb}) preserves the $N$-photon subspaces 
and the steady state in each one of them is a completely random state
\begin{equation}
  \op{\varrho} (t \rightarrow \infty ) =
  \frac{1}{N + 1} \, \op{\openone} \, ,
\end{equation}
which is an important benefit of our approach. The depolarization
rates (\ref{gam}) are very small when compared with other typical
system parameters (observe the dependence on $\Delta_{a}^{-2}$), 
in agreement with experimental observations.

The terms $\mathcal{L} [ \op{S}_{\pm} ]$ describe depolarization in
each invariant subspace, meanwhile the action of $\mathcal{L} [
\op{S}_{0}]$ therein is trivial.  Nevertheless,  $\mathcal{L} [
\op{S}_{0} ]$ is responsible for the relative phase decay between 
blocks of the density matrix corresponding to different excitation 
numbers.

As an illustrative example, we focus on one-photon states.  We are
then within a two-dimensional subspace and the density matrix can 
be expressed as
\begin{equation}
  \op{\varrho} = \frac{1}{2}
  \left (
    \begin{array}{cc}
       1 + z & x - i y \\
      x + i y & 1- z
    \end{array}
  \right )
  =
  \frac{1}{2}
  ( \openone + \mathbf{r} \cdot \op{\bm{\sigma}} ) \, ,
\end{equation}
where $\mathbf{r} = 2 \Tr\ ( \op{\varrho} \op{\bm{\sigma}} )$,
$\op{\bm{\sigma}}$ being the Pauli matrices. The vector
$\mathbf{r}$ satisfies $|\mathbf{r} | \le 1$, with the
equality for pure states (that are all them SU(2) coherent
states~\cite{Perelomov1986}).

In this subspace the master equation~(\ref{mesb}) can be recast in
terms of $\mathbf{r} (t)$ and one obtains the solution as
\begin{eqnarray}
  x (t) & = & x (0) e^{- \gamma t } \, ,  \nonumber \\
  y (t) & = & y (0) e^{- \gamma t} \, , \\
  z (t) & = & z (0) e^{-2 \gamma t} \, .  \nonumber
\end{eqnarray}
If we apply the definition (\ref{doP}) to this case, we get
\begin{equation}
  \mathbb{P} (t) = \frac{1}{2} | \mathbf{r} (t) | \, e^{- \gamma t} \, .
\end{equation}
This degree tends then to zero with a typical time scale $\gamma^{-1}$,
which, as commented before, is exceedingly large.

\section{The quasiclassical regime}

The first term in the master equation~(\ref{mesb}) produces only free
evolution and can be dropped in what follows. In consequence, we
consider the depolarizing equation
\begin{equation}
  \label{mec}
  \dot{\op{\varrho}} = \Gamma \, \mathcal{L} [\op{S}_{0}] \, 
  \op{\varrho}
  + \gamma \, \mathcal{L}   [\op{S}_{+}] \, \op{\varrho}
  + \gamma \, \mathcal{L} [\op{S}_{-}]  \, \op{\varrho} \, ,
\end{equation}
where, for the sake of generality, we have considered different
decoherence rates for $\op{S}_{0}$ and $\op{S}_{\pm}$. To work out the
solutions of (\ref{mec}) we resort to phase-space methods, since they
provide the most suitable approach to study the quasiclassical
limit.  According to the discussion in Sect.~\ref{su2}, one
could think in using quasiprobability distributions on the
sphere~\cite{Stratonovich1956,Berezin1975,Agarwal1981,Klimov2002}.
However, although naturally related to the SU(2) group, they
are not so widely employed by the quantum-optics community, so
we prefer to take a more standard approach. Therefore, we shall use
the $s$-parametrized quasiprobability distributions introduced by
Cahill and Glauber~\cite{Cahill1969}, which for our two-mode fields
read as
\begin{eqnarray}
  \label{Ws}
  W^{(s)} (\alpha_{+}, \alpha_{-}) & = & \frac{1}{\pi^2}
  \int d^2\beta_{+} d^2\beta_{-} \, 
  \chi^{(s)} (\beta_{+}, \beta_{-}) \nonumber \\
  & \times & \exp \left [ \sum_{\lambda = \pm}
    (\alpha_{\lambda} \beta_{\lambda}^\ast - 
    \alpha_{\lambda}^\ast \beta_{\lambda})
  \right ]  \, ,
\end{eqnarray}
where $\chi^{(s)} (\beta_{+}, \beta_{-})$ is the $s$-ordered characteristic
function
\begin{equation}
  \chi^{(s)} (\beta_{+}, \beta_{-}) = \Tr [\op{\varrho} \,
  \op{D}^{(s)} (\beta_{+}, \beta_{-})] \, ,
\end{equation}
and $\op{D}^{(s)} (\beta_{+}, \beta_{-})$ is the two-mode displacement
operator
\begin{equation}
  \op{D}^{(s)} (\beta_{+}, \beta_{-}) =  
  \prod_{\lambda = \pm} e^{s |\beta_\lambda|^2/2}
  \exp ( \beta_{\lambda} \op{a}_{\lambda}^\dagger - 
  \beta_{\lambda}^\ast \op{a}_{\lambda} ) \, .
\end{equation}
In the three special cases $s=0, +1,$ and $-1$, one easily recognizes 
the Husimi $Q$, the Wigner $W$, and the Glauber-Sudarshan $P$ functions,
respectively. In terms of $W^{(s)}$, Eq.~(\ref{mec}) can be transformed
into a differential equation using the following rules of
mapping~\cite{Perina1991}
\begin{eqnarray}
  \label{mapp}
  \op{a}_{\lambda} \op{\varrho} & \mapsto & \left ( 
    \alpha_{\lambda}  + \frac{1-s}{2} 
    \frac{\partial}{\partial \alpha_{\lambda} ^\ast} \right ) 
  W^{(s)} \, ,
  \nonumber \\
  \op{a}_{\lambda}^{\dagger} \op{\varrho} & \mapsto & \left ( 
    \alpha_{\lambda}^\ast - \frac{1+s}{2} 
    \frac{\partial }{\partial \alpha_{\lambda} } \right) 
  W^{(s)} \, ,
  \nonumber \\
  \op{\varrho} \op{a}_{\lambda}  & \mapsto & \left ( 
    \alpha_{\lambda}  - \frac{1+s}{2} 
    \frac{\partial}{\partial \alpha_{\lambda} ^\ast} \right) 
  W^{(s)} \, ,
  \nonumber \\
  \op{\varrho} \op{a}_{\lambda}^{\dagger} & \mapsto & \left ( 
    \alpha_{\lambda}^\ast + \frac{1-s}{2} 
    \frac{\partial }{\partial \alpha_{\lambda}} \right) 
  W^{(s)} \, .
\end{eqnarray}
For definiteness, we choose to work with the $P$ function, which will
facilitate the following calculations. Applying the rules (\ref{mapp})
for each mode, we get that Eq.~(\ref{mec}) can be equivalently written
as
\begin{widetext}
  \begin{eqnarray}
    \label{put}
    \dot{P}( \alpha_{+} , \alpha_{-} ) & = & - \Gamma  \left (
      \alpha_{+}  \, \partial_{\alpha_{+}} -
      \alpha_{+}^\ast \, \partial_{\alpha_{+}^\ast} -
      \alpha_{-} \, \partial_{\alpha_{-} } +
      \alpha_{-}^\ast \, \partial_{\alpha_{-}^\ast} \right)^{2} \,
    P(\alpha_{+} , \alpha_{-}) \nonumber \\
    & + & 2 \gamma \left [
      | \alpha_{-} |^{2} \, \partial_{\alpha_{+}} \partial_{\alpha_{+}^\ast} +
      | \alpha_{+} |^{2} \, \partial_{\alpha_{-}} \partial_{\alpha_{-}^\ast} -
      \alpha_{+} \alpha_{-} \, \partial_{\alpha_{+}} \partial_{\alpha_{-}} -
      \alpha_{+}^\ast \alpha_{-}^\ast \, \partial_{\alpha_{+}^\ast}
      \partial_{\alpha_{-}^\ast} \right . \nonumber \\
    & - & \left . \frac{1}{2} \left ( \alpha_{+} \, \partial_{\alpha_{+}} +
        \alpha_{+}^\ast \, \partial_{\alpha_{+}^\ast} +
        \alpha_{-} \, \partial_{\alpha_{-} } +
        \alpha_{-}^\ast \, \partial_{\alpha_{-}^\ast} \right ) \right ] \,
    P(\alpha_{+} ,\alpha_{-} ) \, .
  \end{eqnarray}
\end{widetext}
At first glance, this equation looks very intricate to be of any practical
value. However, let us  introduce the following differential operators
\begin{eqnarray}
  \label{diffrea}
  \op{\mathcal{S}}_{0} & = &  
  \alpha _{-} \, \partial_{\alpha_{-}} 
  -  \alpha_{-}^{\ast} \, \partial_{\alpha_{-}^{\ast}}
  +  \alpha_{+} \, \partial_{\alpha _{+}}
  - \alpha_{+}^{\ast} \, \partial_{\alpha_{+}^{\ast}}  \, , 
  \nonumber \\
  \op{\mathcal{S}}_{\pm} & = & 
  \alpha_{\mp} \, \partial_{\alpha_{\pm}} 
  - \alpha_{\pm}^{\ast} \, \partial_{\alpha_{\mp}^{\ast}}  \, ,  \\
  \op{\mathcal{S}}_{3} & = &  
  \alpha _{-} \, \partial_{\alpha_{-}} 
  -  \alpha_{-}^{\ast} \, \partial_{\alpha_{-}^{\ast}}
  - \alpha _{+}\,\partial _{\alpha _{+}}
  + \alpha _{+}^{\ast} \, \partial_{\alpha _{+}^{\ast }}  \, ,  
  \nonumber
  \label{L12} 
\end{eqnarray}
which are nothing but a differential realization of the Stokes
operators (i.e., of the u(2) algebra). It turns out that
Eq.~(\ref{put}) reduces to the very simple factorized form
\begin{equation}
  \label{med}
  \dot{P} (\alpha_{+}, \alpha_{-}) = - 4 
  \left[
    \Gamma \op{\mathcal{S}}_{0}^{2} +  \gamma ( 
    \op{\mathcal{S}}_{+} \op{\mathcal{S}}_{-} +
    \op{\mathcal{S}}_{-} \op{\mathcal{S}}_{+} )
  \right] \, P(\alpha_{+}, \alpha_{-}) \, .
\end{equation}
But $\op{\mathcal{S}}^{2} = \op{\mathcal{S}}_{3}^{2} + 
( \op{\mathcal{S}}_{+} \op{\mathcal{S}}_{-} + \op{\mathcal{S}}_{-}
\op{\mathcal{S}}_{+} )/2$ is the Casimir operator for su(2), which
reduces to $S (S + 1) \op{\openone}$ in each invariant subspace 
($S$ taking only integer or half-odd integer values).  Then, this 
master equation has the formal solution
\begin{eqnarray}
  P (\alpha_{+},\alpha_{-}|t ) & = & \exp \{- 4 [
  \Gamma \op{\mathcal{S}}_{0}^{2} +  2 \gamma (
  \op{\mathcal{S}}^{2} - \op{\mathcal{S}}_{3}^{2})] t \} 
  \nonumber \\
  & \times & P(\alpha_{+},\alpha_{-}|t=0) \, .
\end{eqnarray}
Because of the properties of the $P$ representation, the average value
of any observable $\op{O} (t)$ can be recast as
\begin{eqnarray}
  \label{vm}
  \langle \op{O} (t) \rangle  & = & \int d^{2}\alpha_{+} d^{2}\alpha_{-} \, 
  P(\alpha_{+},\alpha_{-}|t=0)  \nonumber \\
  & \times & \exp \{- 4 [ \Gamma \op{\mathcal{S}}_{0}^{2} + 
  2  \gamma (\op{\mathcal{S}}^{2} - \op{\mathcal{S}}_{3}^{2})] \, 
  \langle \alpha_{+}\alpha_{-}|\op{O}|\alpha_{+}\alpha_{-}\rangle \, ,
\nonumber \\
  &&
\end{eqnarray}
where by $\op{O}$ we mean the operator at $t=0$. The problem for 
a practical application of this equation is that we need to give 
meaning to the action of the exponential. To this end,  we first 
introduce the parametrization
\begin{eqnarray}
  \label{cs2}
  \alpha_{+} & = & r \, e^{-i(\phi +\psi )/2} \cos (\theta /2 ) \, ,  
  \nonumber \\
  && \\
  \alpha_{-} & = & r \, e^{ i ( \phi -\psi )/2 }\sin (\theta /2) \, ,  
  \nonumber
\end{eqnarray}
where $r^{2} = | \alpha_{+} |^{2} + | \alpha_{-} |^{2}$ is a radial
variable related with the global intensity. The angles $\theta $ and
$\phi $ can be interpreted as the polar and azimuthal angles,
respectively, on the Poincar\'{e} sphere, while $\psi$ is a common
phase.  These coordinates reflect the fact that polarization needs
only three independent quantities to be fully characterized: the
amplitudes of each mode and the relative phase between them.

If we concentrate in the evolution of the components of the Bloch 
vector $\langle \op{\mathbf{S}} \rangle$, we have
that
\begin{eqnarray}
  \langle \alpha_{+}, \alpha_{-}| \op{S}_{3}  |\alpha_{+}, \alpha_{-} \rangle
  & = & r^{2} \cos \theta =  r^{2} D_{00}^{1}(\phi ,\theta ,\psi ) \, , 
  \nonumber \\
  & & \\
  \langle \alpha_{+}, \alpha_{-}| \op{S}_{\pm} |\alpha_{+}, \alpha_{-} \rangle
  & = & r^{2} e^{\mp i\phi } \sin \theta = 
  \sqrt{2}r^{2} D_{\pm1 0}^{1}(\phi ,\theta ,\psi ) \, ,
  \nonumber
\end{eqnarray}
where $D_{m m^\prime}^{S}$ are Wigner
$D$-functions~\cite{Varshalovich1988}, which constitute an orthogonal
basis of complex-valued functions defined on the irreducible carrier
subspaces of su(2), so that the action of the operators
(\ref{diffrea}) on them is standard.  Then a direct application of
(\ref{vm}) yields
\begin{eqnarray}
  \langle \op{S}_{3}(t) \rangle  & = & \exp (-8 \gamma t ) \,
 \langle \op{S}_{3}(0) \rangle \, ,  \nonumber \\
  & & \\
  \langle \op{S}_{\pm }(t) \rangle  & = & \exp  [ - (2 \gamma +
 \Gamma ) t ] \, \langle \op{S}_{\pm }(0)\rangle \, . \nonumber
\end{eqnarray}

The procedure can be applied much in the same way for the field amplitudes;
we merely quote the simplest results:
\begin{eqnarray}
  \langle \op{a}_{\pm} (t) \rangle  & = & \exp [ - (\gamma + \Gamma /4) t ]
  \langle \op{a}_{\pm} (0) \rangle \, , 
  \nonumber \\
  & & \\
  \langle \op{a}_{\pm}^{2} (t) \rangle  & = & \exp [-2 (\gamma + \Gamma) t ]
  \langle  \op{a}_{\pm}^{2}(0) \rangle \, .
  \nonumber
\end{eqnarray}
The rest of the moments can be obtained in similar fashion.

On the other hand, in terms of (\ref{cs2}) the operators
$\op{\mathcal{S}}_{0}$ and $\op{\mathcal{S}}_{3}$ take the suggestive
form~\cite{Varshalovich1988}
\begin{equation}
  \op{\mathcal{S}}_{0}  =  - i \partial_{\psi} \, ,
  \qquad
  \op{\mathcal{S}}_{3} = - i \partial_{\phi } \, ,  
\end{equation}
and the exact solution of (\ref{med}) can be expressed as
\begin{eqnarray}
  \label{solmed}
  P(\phi , \theta , \psi |t)  & = & \sum_{S}^{\infty }
  \sum_{m,m^{\prime} = - S}^{S} c_{mm^{\prime}}^{S} \,
  D_{m m^{\prime}}^{S}(\phi , \theta, \psi )  \nonumber \\
  & \times &\exp \{ [-8 \gamma \, S ( S + 1) +  8  \gamma m^{2} - 
  4 \Gamma {m^{\prime}}^{2}] t \} \, , \nonumber \\
\end{eqnarray}
where the coefficients $c_{nn^{\prime }}^{S}$ are determined by
the initial state:
\begin{eqnarray}
  \label{coeff}
  c_{m m^{\prime}}^{S} & = & \frac{8\pi^{2}}{2 S + 1}
  \int_{0}^{2\pi} \!\!\! d\psi 
  \int_{0}^{\pi} \!\!\! d\theta \, \sin \theta
  \int_{0}^{2\pi} \!\!\! d\phi \,
  D_{m m^{\prime}}^{S \, {}^\ast} (\phi , \theta , \psi )  \nonumber \\
  & \times & P (\phi , \theta , \psi |t = 0) \, .
\end{eqnarray}
We can  proceed to integrate over the physically irrelevant global 
phase $\psi$, obtaining in this way the $P$ function over the 
unit sphere. The result is
\begin{eqnarray}
  P (\phi ,\theta |t) & = & \sum_{S}^{\infty} 
  \sqrt{\frac{4\pi}{2 S +1}}
  \sum_{m = - S }^{S}  
  c_{m 0}^{S} \, Y_{S m} (\theta , \phi )
  \nonumber \\
  & \times & \exp \{-8  \gamma [S ( S + 1)  - m^{2}] t \} \, ,
\end{eqnarray}
where $ Y_{S m} (\theta , \phi )$ are spherical harmonics. This is
a closed formula that can be considered as our major result: it 
allows to trace the depolarizing dynamics on the Poincar\'e sphere
for arbitrary states. For example, for (quadrature) coherent states, 
we have
\begin{equation}
  P (\alpha_{+}, \alpha_{-} |t=0) =  \delta^{2} (\alpha_{+} - r_{0})
  \delta^{2} ( \alpha_{-}- r_{0}) \, ,
\end{equation}
where we have assumed that both modes have the same real amplitude 
$r_0$. We then get
\begin{equation}
  \label{Pdel}
  P( \phi ,\theta ,\psi |t=0) = 8 \sqrt{2}  \, 
  \frac{\delta (r-r_{0})}{r^{3}}  \, 
  \delta ( \psi ) \, \delta ( \phi ) \, \delta ( \theta - \pi/2 ) \, ,
\end{equation}
and so 
\begin{equation}
  c_{m m^\prime}^{S} = \frac{64 \sqrt{2} \pi^{2}}{2 S + 1}
  D_{m m^\prime}^{S \, {}^\ast} (0, \pi/2, 0) \frac{\delta (r - r_{0})}
  {r^{3}} \, .
\end{equation}
From here we can immediately compute the dynamics of this state.

In view of the delta functions appearing in (\ref{Pdel}), one can
argue  that the $P$ representation results in a distribution with
singularities. We can look instead at smoother quasidistributions, 
such as, e.g., the $Q$ function. Curiously enough, when we repeat 
the calculations, we end up with the fact that the evolution equation 
for the $Q$ function is exactly the same as for the $P$  function. 
In consequence, the  time evolution of the $Q$ function can be 
expressed as in Eq.~(\ref{solmed}) [obviously, the coefficients 
$c_{m m^\prime}^{S}$ are determined now as in (\ref{coeff}), but with 
$P$ replaced by $Q$].

This can be used to determine any function of $Q$. In fact,
it has been argued~\cite{Luis2002} that, for a given field, the
distance between its $Q$ function and the $Q$ function for unpolarized
light can be taken as a proper degree of polarization in phase
space: this distance is, except from constant factors, proportional
to $Q^{2}$. Therefore, we can take as an unnormalized depolarization
 measure
\begin{equation}
  \mathbb{D} (t) = 
  \int_{0}^{2\pi} \!\!\! d\psi 
  \int_{0}^{\pi} \!\!\! d\theta \, \sin \theta
  \int_{0}^{2\pi} \!\!\! d\phi \, 
  Q^{2}(\phi ,\vartheta ,\psi |t),
\end{equation}
and then we obtain
\begin{eqnarray}
  \mathbb{D} (t) & = & \sum_{S}^{\infty} \frac{4\pi}{2 S +1} 
  \sum_{m, m^{\prime} = -S}^{S} |c_{m m^{\prime}}^{S}|^{2} \nonumber \\
  & \times & \exp [-8 ( 2 \gamma S (S + 1) - 2 \gamma m^{2} + 
  \Gamma {{m^\prime}}^{2} ) ] .
\end{eqnarray}
Again we can ascertain the corresponding time evolution for
arbitrary states.

\section{Concluding remarks}

In this paper we have reported a comprehensive, simple theory of
quantum light depolarization. In our model the field couples
dispersively to a randomly distributed atomic reservoir, and the
resulting master equation has unique properties that we have explored
in detail.

We have solved this master equation resorting to a simple phase-space
formalism for polarization on the Poincar\'e sphere, based on
$s$-ordered quasidistributions for the two basic polarization
modes. These results may have interesting consequences to implement
experimental procedures for determining polarization properties.


\end{document}